\documentclass[twocolumn,showpacs,floats,floatfix,superscriptaddress,aps,pra]{revtex4}

\usepackage{amsfonts}
\usepackage{amssymb}
\usepackage{amsmath}
\usepackage{calc}
\usepackage[dvips]{graphicx}
\usepackage{bm}

\def\be{\begin{equation}}
\def\ee{\end{equation}}
\def\bea{\begin{eqnarray}}
\def\eea{\end{eqnarray}}
\def\bse{\begin{subequations}}
\def\ese{\end{subequations}}

\def\1{\mathbf{1}}

\begin{document}

\title{Dynamical control and novel quantum phases in impurity doped linear ion crystals}

\date{\today}
\author{Peter A. Ivanov}
\affiliation{Institut f\"{u}r Quanteninformationsverarbeitung,
Universit\"{a}t Ulm, Albert-Einstein-Allee 11, 89081 Ulm, Germany}
\affiliation{Department of Physics, Sofia University, James
Bourchier 5 blvd, 1164 Sofia, Bulgaria}
\author{Nikolay V. Vitanov}
\affiliation{Department of Physics, Sofia University, James
Bourchier 5 blvd, 1164 Sofia, Bulgaria}
\author{Kilian Singer}
\affiliation{Institut f\"{u}r Quanteninformationsverarbeitung,
Universit\"{a}t Ulm, Albert-Einstein-Allee 11, 89081 Ulm, Germany}
\author{Ferdinand Schmidt-Kaler}
\affiliation{Institut f\"{u}r Quanteninformationsverarbeitung,
Universit\"{a}t Ulm, Albert-Einstein-Allee 11, 89081 Ulm, Germany}

\begin{abstract}
We explore the behavior of the phonon number distribution in an
heterogeneous linear ion crystal. The presence of ion species with
different masses changes dramatically the transverse energy
spectrum, in such a way that two eigenfrequencies become
non-analytic functions of the mass ratio in the form of a sharp
cusp. This non-analyticity induces a quantum phase transition
between condensed and conducting phase of the transverse local
phonons. In order to continuously vary the mass ratio we
adiabatically modify a locally applied laser field, exerting
optical dipole forces which reduces the effective mass.
\end{abstract}

\pacs{03.67.Ac, 37.10.Ty, 64.70.Tg, 73.43.Nq} \maketitle

\section{Introduction}

Trapped ions are one of the most attractive physical systems for
implementing quantum computation \cite{Haffner} and quantum
simulation \cite{MJ}. The ability to control and measure the
internal and external degrees of freedom with high accuracy allows
for experimental implementations of various quantum gates
\cite{FSK,DL,TM,KK} and quantum computing protocols
\cite{SG,JC,KB}. Trapped ions represent a convenient system for
the simulation of many-body effects, such as quantum phase
transitions in spin systems \cite{DC,AF}, atoms in optical
lattices \cite{DC2} and coupled cavity arrays \cite{PAI}.
Recently, a simulation of the Dirac equation for a free spin-1/2
particle \cite{LL} has been successfully performed with a single
trapped ion \cite{CR}.

\begin{figure}[tb]
\includegraphics[angle=0,width=50mm]{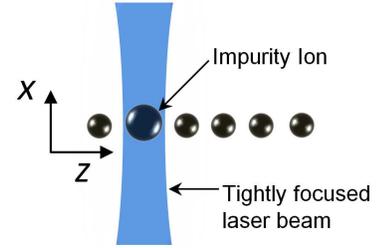}
\caption{(color online). Impurity-doped linear ion crystal
consisting of ions with mass $m$ and an impurity ion with mass
$M$. The mass ratio $\mu=M/m$ determines the two quantum phases.
Applying a tightly focused laser field to the impurity ion in the
transverse $x$ direction makes it possible to continuously vary
the effective mass ratio. Increasing the magnitude of the light
field leads to a decrease of the effective mass.} \label{fig1}
\end{figure}

While small, homogeneous ion crystals have been extensively
studied, the novelty of this paper is elucidating the richness of
quantum phases in ion crystals doped with a second ion species.
Depending on the mass ratio, the crystal allows for the
observation of two quantum phases of the transverse \emph{local}
phonon number distribution. For a heavy impurity ion all phonons
are condensed by the latter and the variance of the transverse
local phonon distribution is zero. At the same time zero average
phonons with vanishing variance and correlation are observed at
all other sites. In the following this state is referred to as the
quantum \emph{condensed} phase. In the opposite case, if a lighter
impurity ion is added, the system is in a quantum
\emph{conducting} phase \cite{DC2} where the phonons are
redistributed among all ions showing non-zero variance and
correlation, while the average phonons of the impurity ion is
significantly reduced. There are two ways to vary the mass ratio
in heterogeneous ion crystal. The first one is by doping a
mono-species ion crystal with one impurity ion, a technique which
has been applied for ion frequency standards \cite{TR},
sympathetic cooling of molecular ions \cite{MBarrett}, and for
deterministic ion implantation \cite{WS}. The second scheme
overcomes this discrete, stepwise variation and uses optical
dipole forces by non-resonant tightly focused laser beams
\cite{TS} (Fig. \ref{fig1}). It is still advantageous to use
heavier doping ions because the dipole force can only reduce the
effective mass but to observe the phase transition the critical
point of unity mass ratio has to be crossed.
Current ion trap technology allows to implement all crucial
elements of the proposal: preparation of the initial state,
laser-ion interaction driving the dynamics required for the
quantum phase transition, and readout of the phonon distribution.

\section{Transverse Energy Spectrum}

In the following we first consider a harmonically confined ion
crystal with $N-1$ ions of mass $m$ and one impurity ion of mass
$M$ at position $j_{M}$ without any focused laser field. If the
radial confinement is stronger than the axial one then ions are
arranged in a linear ion crystal along the axial $z$ axis and
occupy equilibrium positions $z_{i}^{0}$. Since the axial trap
potential is independent of the mass, the equilibrium positions
$z_{i}^{0}=lu_{i}$ of the ions are independent of the composition
of the ion crystal. Here $l$ is the natural length scale and
$u_{i}$ are dimensionless equilibrium positions \cite{James}. The
transverse, dynamic Paul confinement along $x$ and $y$ directions
is generated by an applied radio frequency (RF) quadrupole field.
In the following we only consider one transverse direction $x$.
The oscillation frequency in $x$ direction is
$\omega_{x}^{0}=|e|U_{\text{RF}}c_{x}/(\sqrt{2}m\Omega_{\text{RF}})$,
where $U_{\text{RF}}$ is the amplitude of the RF voltage with
frequency $\Omega_{\text{RF}}$, $e$ is the electron charge and
$c_{x}$ is a geometry factor. The transverse oscillation frequency
is additionally reduced by the axial trap frequency $\omega_{z}$
according to $\omega_{x}=\omega_{x}^{0}(1-\alpha^{2}/2)^{1/2}$, in
lowest-order approximation, where
$\alpha=\omega_{z}/\omega_{x}^{0}$ \cite{LBMW}. For a small
displacement the motional degrees of freedom in $x$, $y$ and $z$
direction are decoupled. The normal mode eigenfrequencies
$\omega_{k}=\omega_{x}^{0}\sqrt{\lambda_{k}}$ and eigenvectors
$b_{j}^{k}$ are obtained from the diagonalization of a matrix
$B_{ij}$ with $\sum_{i}B_{ij}b_{i}^{k}=\lambda_{k}b_{j}^{k}$,
where $\omega_{x}^{0}$ is the transverse frequency of a single ion
with mass $m$. The real and symmetric $N$-dimensional matrix
$B_{ij}$ is determined by the harmonic expansion of the external
trapping potential and the Coulomb interaction between the ions;
it reads \cite{KZ}
\begin{equation} \label{B}
B_{ij}=\left\{\begin{array}{c}
1-\dfrac{\alpha^{2}}{2}-\alpha ^{2} \sum_{\substack{ p=1 \\ p\neq j}}^{N}\dfrac{1}{\left\vert u_{j}-u_{p}\right\vert ^{3}},\ (i=j,\ j\neq j_{M}), \\
\dfrac{1}{\mu ^{2}}-\dfrac{\alpha^{2}}{2\mu}-\dfrac{\alpha ^{2}}{\mu}\sum_{\substack{ p=1 \\ p\neq j}}^{N}\dfrac{1}{\left\vert u_{j}-u_{p}\right\vert ^{3}},\ (i=j=j_{M}),\\
\dfrac{\alpha ^{2}}{|u_{i}-u_{j}|^{3}},\ (i\neq j\neq j_{M}), \\
\dfrac{\alpha ^{2}}{\sqrt{\mu}|u_{i}-u_{j}|^{3}},\ (i\neq
j=j_{M}),
\end{array}\right.
\end{equation}
where $\mu=M/m$ denotes the mass ratio of both ion species. By
expressing the position and momentum operators in terms of normal
phonon modes
\begin{eqnarray}\label{XP}
\hat{{x}}_{j}=\sum_{k}b_{j}^{k}\hat{X_{k}},\hspace{4mm} \hat{{p}}_{j}=\sum_{k}b_{j}^{k}\hat{P_{k}}, \notag \\
\hat{{x}}_{j_{M}}=\frac{1}{\sqrt{\mu}}\sum_{k}b_{j_{M}}^{k}\hat{X_{k}},\hspace{4mm}
\hat{{p}}_{j_{M}}=\sqrt{\mu}\sum_{k}b_{j_{M}}^{k}\hat{P_{k}},
\end{eqnarray}
the Hamiltonian takes the form
$\hat{H}_{0}=\sum_{k}\hbar\omega_{k}(\hat{N}_{k}+1/2)$. Here
$\hat{X_{k}}=\sqrt{\hbar /2
m\omega_{k}}(\hat{a}_{k}^{\dagger}+\hat{a}_{k})$ is the normal
mode position operator and $\hat{P_{k}}=\text{i}\sqrt{\hbar m
\omega_{k}/2}(\hat{a}_{k}^{\dagger}-\hat{a}_{k})$ is the normal
mode momentum operator, while $\hat{a}_{k}^{\dagger}$ and
$\hat{a}_{k}$ are the phonon creation and annihilation operators
of the $k$th collective phonon mode and
$\hat{N}_{k}=\hat{a}_{k}^{\dagger} \hat{a}_{k}$ is the respective
\emph{collective} phonon number operator.

\begin{figure}[tb]
\includegraphics[angle=0,width=85mm]{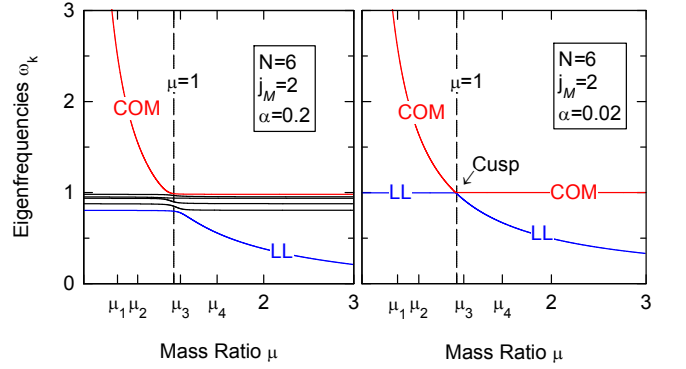}
\caption{(color online). Normalized transverse eigenfrequencies
$\omega_{k}/\omega_{x}^{0}=\sqrt{\lambda_{k}}$ when $\mu$ is
varied. The highest energy center-of-mass (COM) and the
lowest-lying (LL) modes are shown. Different ion species are
depicted as follows:
$\mu_{1}=$$^{9}\textrm{Be}^{+}/^{24}\textrm{Mg}^{+}$,
$\mu_{2}=$$^{24}\textrm{Mg}^{+}/^{40}\textrm{Ca}^{+}$,
$\mu_{3}=$$^{40}\textrm{Ca}^{+}/^{43}\textrm{Ca}^{+}$ and
$\mu_{4}=$$^{40}\textrm{Ca}^{+}/^{27}\textrm{Al}^{+}$. b)
Decreasing $\alpha$ the LL and COM modes develop a cusp at
$\mu=1$.} \label{fig2}
\end{figure}

In order to visualize the transverse mode we plot in
Fig.~\ref{fig2} the eigenfrequencies $\omega_{k}$ for a string of
six ions, with an impurity ion at position $j_{M}=2$ when the mass
ratio is varied. As $\mu$ increases beyond unity, the frequency
spacing between the lowest-lying energy mode (LL) with frequency
$\omega_N$ and all other modes increases, while for $\mu<1$ it
tends towards degeneracy. In the transverse direction the highest
energy mode, the center-of-mass mode (COM) with frequency
$\omega_1$, is nearly degenerate for $\mu>1$. When $\mu$ decreases
below the value $\mu=1$ the frequency spacing between COM mode and
all other modes increases. In real experiments the mass ratio can
only be changed to discrete values as shown in Fig. \ref{fig2}.
However, later in the paper we show how to overcome this
limitation by optical forces.

As $\alpha$ is decreased the gap at the avoided level-crossing
($\mu=1$) between the modes vanishes. 
Moreover, LL and COM modes develop a cusp at $\mu=1$. For
sufficiently small $\alpha\ll 1$ the off-diagonal elements of
$B_{ij}$ (\ref{B}) can be neglected and the the eigenfrequencies
$\omega_{k}$ behave as follows
\begin{equation}
\left.\left.
\begin{array}{c}
\omega_{1}\rightarrow \dfrac{\omega_{x}}{\mu} \notag \\
\omega_{n\neq 1}\rightarrow \omega_{x}
\end{array}
\right\} \text{ for } \mu<1,\quad
\begin{array}{c}
\omega_{n\neq N}\rightarrow \omega_{x} \notag \\
\omega_{N}\rightarrow \dfrac{\omega_{x}}{\mu}
\end{array}
\right\} \text{ for } \mu>1.\label{FR}
\end{equation}
This implies that the  COM frequency $\omega_{1}$ and the  LL
frequency $\omega_{N}$, respectively, have discontinuous
derivatives at $\mu=1$. Therefore, the non-analyticity of the
ground state energy indicates a quantum phase transition
\cite{SS}. We will exemplify the properties of the two quantum
phases by calculating the phonon number distribution, its variance
and correlation. These quantities are all experimentally
accessible by laser-spectroscopy measurements.

\section{Quantum Phases}

The characteristic feature of the quantum phase transition is
expressed clearly by the behavior of the \emph{local} phonon
number operator
\begin{eqnarray}
\hat{n}_{j}=\frac{m\omega_{x}^{0}(m)}{2\hbar}\hat{{x}}_{j}^{2}+\frac{\hat{{p}}_{j}^{2}}{2\hbar
m\omega_{x}^{0}(m)}-\frac{1}{2},  \notag \\
\hat{n}_{j_{M}}=\frac{M\omega_{x}^{0}(M)}{2\hbar}\hat{{x}}_{j_{M}}^{2}+\frac{\hat{{p}}_{j_{M}}^{2}}{2\hbar
M\omega_{x}^{0}(M)}-\frac{1}{2}.
\end{eqnarray}
Initially the mixed ion crystal is prepared by laser-ion
interactions in the state with $n$ phonons in the LL collective
mode and the other modes are cooled to the ground state, achieving
the state $|\Psi\rangle=|00\ldots n\rangle$. We find that the
average local phonons per site (per ion) $\langle
\hat{n}_{j}\rangle\equiv\langle \Psi|\hat{n}_{j}|\Psi\rangle$ is
given by
\begin{equation}
\langle \hat{n}_{j} \rangle
=\frac{n(b_{j}^{N})^{2}}{2}\left(\frac{\omega_{N}}{\omega_{x}^{0}}+\frac{\omega_{x}^{0}}{\omega_{N}}\right)+
\frac{1}{4}\sum_{k=1}^{N}(b_{j}^{k})^{2}\left(\frac{\omega_{k}}{\omega_{x}^{0}}+\frac{\omega_{x}^{0}}{\omega_{k}}\right)-\frac{1}{2}.\label{PN}
\end{equation}
First we will discuses the properties of Eq. (\ref{PN}) for
$\mu=1$. The results implies that in general $\sum_{j}\langle
n_{j}\rangle\geq n$. The inequality does not violate the
conservation of energy since the sum of the local phonons energies
is decreased due to Coulomb interaction between the ions. However,
if $\alpha$ is kept sufficiently small ($\alpha\ll 1$) the
transversal vibrational modes are approximately degenerate
$\omega_{k}\approx \omega_{x}^{0}$ and the local phonons per site
are $\langle n_{j}\rangle=n(b_{j}^{N})^{2}$ so that
$\sum_{j}\langle n_{j}\rangle=n$. In this regime where the Coulomb
energy is much weaker than the trapping energy the linear ion
crystal is in the quantum conducting (superfluid) phase and the
phonon number $n$ is conserved \cite{DC2}. Now we will discuss the
properties of Eq. (\ref{PN}) in the two limits $\mu>1$ and
$\mu<1$, respectively. As can be seen from Fig. \ref{fig2}b for
$\mu>1$ and $\alpha\ll 1$ all modes are degenerate except
$\omega_{N}=\omega_{x}^{0}/\mu$. Moreover, we find that
$b_{j_{M}}^{N}=1$ and $b_{j_{M}}^{k\neq N}=0$ using Eq.~(\ref{B}).
This implies that the heavy ion is oscillating while the others
are at rest. The average number of transversal phonons is given by
\begin{figure}[tb]
\includegraphics[angle=-90,width=80mm]{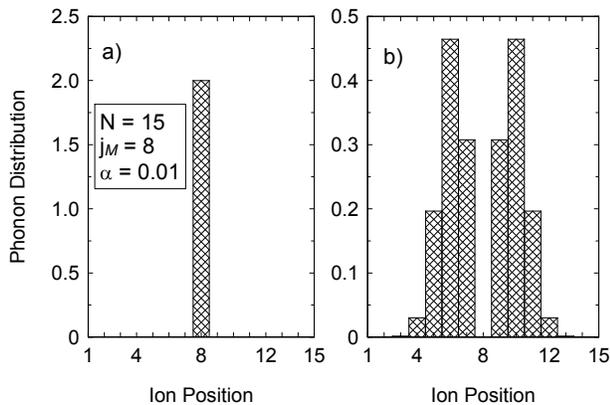}
\caption{Average phonon number distribution as described by Eqs.
(\ref{PN}) and (\ref{VarL}). The impurity ion is placed in the
center of a string of $N=15$ ions, i.e. at position $j_{M}=8$. The
quantum condensed phase is observed in a heterogeneous ion crystal
with fourteen $^{40}$Ca$^+$ ions and one $^{43}$Ca$^+$ impurity
ion, where $n=2$ collective phonons are localized at the impurity
site (a). For $\mu=40/43$ corresponding to fourteen
$^{40}\textrm{Ca}^{+}$ ions and one $^{43}\textrm{Ca}^{+}$
impurity ion the system is in a quantum conducting phase with non
zero phonon distribution (b).} \label{fig3}
\end{figure}
\begin{equation}
\langle \hat{n}_{j\neq j_M} \rangle= 0 ,\hspace{4mm} \langle
\hat{n}_{j_{M}}\rangle= n.\label{H}
\end{equation}
Thus, for a heavy impurity ion, increasing the number of
collective phonons in the LL mode leads to an increase of the
average number of local phonons at the impurity site whereas all
other ions remain in the vibrational ground state. Hence, the
heterogeneous ion crystal is in a quantum condensed phase where
any transversal quantized energy is condensed by the heavy ion.

In the opposite limit of a lighter impurity ion we derive from
Eq.~(\ref{B}) that $b_{j_{M}}^{1}=1$, $b_{j_{M}}^{k\neq 1}=0$.
Because now only the LL modes are excited, the oscillations of the
impurity ion are suppressed and the average number of phonons is
\begin{equation}
\langle \hat{n}_{j\neq j_M}\rangle= n(b_{j}^{N})^{2}, \hspace{4mm}
\langle \hat{n}_{j_{M}}\rangle= 0.\label{L}
\end{equation}
Hence, the lighter impurity ion spreads the phonons to all other
ions, such that the heterogeneous ion crystal is in a quantum
conducting phase. Eq. (\ref{L}) shows that the lighter impurity
ion is in a vibrational ground state and is independent to the
collective phonons $n$. The phonon distribution is shown in
Fig.~\ref{fig3}, where both phases are exemplified for the system
of a mixed $^{43}$Ca$^+$/$^{40}$Ca$^+$ crystal.
\begin{figure}[tb]
\includegraphics[angle=-90,width=65mm]{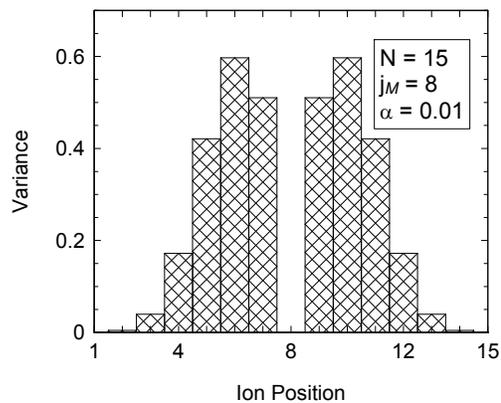}
\caption{The variance of the local phonon number operator as
described by Eq. (\ref{VarL}). The lighter impurity ion is placed
in the center of a string of $N=15$ ions, i.e. at position
$j_{M}=8$. For $\mu=40/43$ corresponding to fourteen
$^{43}\textrm{Ca}^{+}$ ions and one $^{40}\textrm{Ca}^{+}$
impurity ion the system is in a quantum conducting phase. The
phase is characterized with non-zero variance for all sites except
for the impurity ion where the variance is zero.} \label{VRs}
\end{figure}

The different quantum phases become evident also from the variance
of the local phonon number operator $\delta n_{j}=(\langle
\hat{n}_{j}^{2}\rangle)-\langle \hat{n}_{j} \rangle^{2})^{1/2}$.
Again we will start with the case of $\mu=1$, where we obtain that
$\delta n_{j} = \sqrt{n(b_{j}^N)^{2}-n(b_{j}^{N})^{4}}$. However,
for the heavy impurity case ($\mu>1$) we derive $\delta n_{j} = 0$
for all sites. The vanishing phonon variance appears to due the
the localization of the collective phonons at the impunity site.
If the impurity ion is lighter ($\mu<1$) we obtain
\begin{equation}
\delta n_{j\neq j_M} = \sqrt{n(b_{j}^N)^{2}-n(b_{j}^{N})^{4}},
\hspace{4mm} \delta n_{j_{M}} = 0.\label{VarL}
\end{equation}
Now the variance increases with the amount $n$ of phonons in the
LL mode for all ions except for the impurity ion.

Finally, we demonstrate the quantum phases by considering
correlations in the number of phonons $C_{ij}=\langle
\hat{n}_{i}\hat{n}_{j}\rangle-\langle \hat{n}_{i}\rangle
\langle\hat{n}_{j}\rangle$. For $\mu=1$ the ion crystal is in the
quantum conducting phase with correlation
$C_{ij}=-n(b_{i}^{N}b_{j}^{N})^{2}$. Since in the quantum
condensed phase ($\mu>1$) the collective LL phonons are localized
at the impurity ion whereas the remaining ions are in the
vibrational ground state we find that the correlation $C_{ij}$ for
any $i$ and $j$ vanishes. In the quantum conducting phase
($\mu<1$) the lighter impurity ion is in the vibrational ground
state and therefore the correlation $C_{ij_{M}}$ remains zero.
However, for all other ions the correlation is
$C_{ij}=-n(b_{i}^{N}b_{j}^{N})^{2}$.

To summarize, the impurity doped ion crystal allows to distinguish
two quantum phases depends of the mass ratio $\mu$. In the quantum
condensed phase ($\mu>1$) the collective LL phonons are localized
at the impurity ion whereas all other ions remain in the
vibrational ground state. The localization leads to vanishing of
the phonon variance and correlation at all sites. In the quantum
conducting phase ($\mu<1$) the lighter impurity ion is in the
vibrational ground state and the collective LL phonons are
distributed among the remaining ions. This phase is characterized
with non-zero variance and correlation except for the impurity
ion.

In the following, we show how to employ an optical dipole force in
order to change the effective mass of the impurity ion. In this
way $\mu$ can be continuously swept through the quantum phase
transition.

\section{Optical Dipole Force}

We assume that additionally to the trapping potential an optical
dipole force is applied to the impurity ion in the transverse $x$
direction. This force creates a harmonic potential
$V_{\text{d}}=M\omega_s^2\hat{x}_{j_M}^2/2$ with frequency
$\omega_{s}$. The total potential is the sum of the two trapping
potentials and the Coulomb interaction between the ions. The
Hamiltonian for the transverse ion motion is given by \cite{KZ}
\begin{equation}
\hat{H}(t)=\frac{m}{2}\sum_{j=1}^{N}\left(\frac{d}{dt}\hat{\tilde{x}}_{j}\right)^{2}+\frac{m(\omega_{x}^{0})^{2}}{2}\sum_{j,i=1}^{N}\tilde{B}_{ji}\hat{\tilde{x}}_{j}\hat{\tilde{x}}_{i}
\end{equation}
Here for simplicity we normalize the position operators
$\hat{x}_{j}$ as $\hat{\tilde{x}}_{j}=\hat{x}_{j}$ and
$\hat{\tilde{x}}_{j_{M}}=\sqrt{\mu}\hat{x}_{j}$. Using Eq.
(\ref{XP}) the Hamiltonian can be expressed in terms of normal
modes as follows
\begin{equation}\label{Hamiltonian}
\hat{H}(t)= \hat{H}_{0}+\frac{1}{2}\sum_{k,q=1}^{N}\left[m
R_{kq}\hat{X}_{k}\hat{X}_{q} +
S_{kq}(\hat{P}_{k}\hat{X}_{q}-\hat{X}_{k}\hat{P}_{q})\right].
\end{equation}
The new eigenfrequencies $\omega_{k}$ and eigenvectors $b_{j}^{k}$
are obtained by diagonalization of a matrix
$\widetilde{B}_{ij}=B_{ij}+\beta^{2}\delta_{ij_{M}}\delta_{jj_{M}}$,
with $\beta=\omega_{s}/\omega_{x}^{0}$. Due to the time-dependent
optical dipole interaction the normal modes and momenta are
connected by the couplings $R_{kq}=
\sum_{j=1}^{N}\dot{b}_{j}^{k}\dot{b}_{j}^{q}$ and $S_{kq}=
\sum_{j=1}^{N}b_{j}^{k}\dot{b}_{j}^{q}$. Here we denote with dot
the time derivative, occurring because the magnitude of the dipole
force is varied in time.
\begin{figure}[tb]
\includegraphics[angle=0,width=70mm]{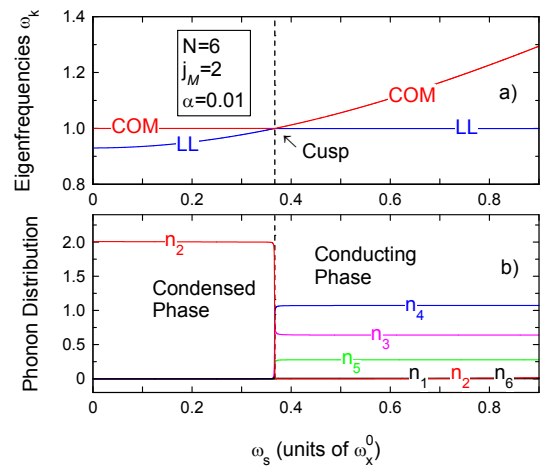}
\caption{(color online). (a) The transverse eigenfrequencies
$\omega_{k}$ for COM and LL modes as a function of the
optical-dipole frequency $\omega_{s}$ for a heterogeneous ion
crystal consisting of  five $^{40}\text{Ca}^{+}$ ions and one
$^{43}\text{Ca}^{+}$ impurity ion at the second position. A
quantum phase transition is observed by adiabatically sweeping
$\omega_s$ to higher values. (b) The local average phonon
distribution for each ion $\langle \hat{n}_{i}\rangle =n_{i}$
($i=1,2\ldots ,6$) as a function of $\omega_{s}$. As the magnitude
of $\omega_{s}$ is changed the mass ratio is reduced and reaches a
critical point $\mu_{\text{eff}}=1$ at which the system undergoes
a quantum phase transition from the quantum condensed phase to the
quantum conducting phase. The quantum phase transition occurs
approximately at $\omega_{s}=0.37 \omega_{x}^{0}$.} \label{fig4}
\end{figure}

The experimental scheme starts by preparing the crystal in state
$|\Psi\rangle = |00\ldots n\rangle$. During an adiabatic increase
of the trap potential created by the optical dipole force the
normal modes become coupled. In order to suppress non-adiabatic
transitions to the other modes we require the adiabatic condition
to be fulfilled at any instant of time, $|S_{kq}| \ll
|\omega_{k}(t)-\omega_{q}(t)|$. Then the couplings $R_{kq}$ and
$S_{kq}$ can be neglected such that the Hamiltonian
\eqref{Hamiltonian} describes $N$ uncoupled harmonic oscillators
with time-dependent frequencies $\omega_{k}$. We assume that the
frequency of the tightly focused laser field is varying as
$\omega_{s}=\omega_{s}^{0}\sqrt{t}$. The energy spectrum of the
linear ion string is now represented by the Demkov-Osherov model
\cite{DO}, which describes the interaction between $N-1$
stationary states and one state varying linearly in time. This
allows us to approximate the adiabatic condition as
$(\omega_{s}^{0}/\omega_{x}^{0})^{2}\lesssim
|\omega_{k}-\omega_{q}|^{2}T $. This condition implies that the
temporal energy change should be smaller than the splitting to the
next level to avoid transitions. Then the ion crystal remains in
the initially prepared state $|\Psi\rangle$ during the adiabatic
change of the optical dipole force. To illustrate the effect of
the optical force on the transverse spectrum we plot in
Fig.~\ref{fig4}a the eigenfrequencies $\omega_{k}$ for the COM and
LL modes as a function of $\omega_{s}$ for five
$^{40}\textrm{Ca}^{+}$ ions and one $^{43}\textrm{Ca}^{+}$
impurity ion at $j_{M}=2$. For sufficiently small $\alpha\ll 1$
the off-diagonal elements of $\tilde{B}_{ij}$ can be neglected and
we obtain that the optical dipole force reduces the mass ratio as
$\mu_{\textrm{eff}}=((\omega_s/\omega_{x}^{0})^{2}+1/\mu^{2})^{-1/2}$.
As the magnitude of $\mu_{\text{eff}}$ is changed, the system
reaches a critical point $\mu=1$ for which the system undergoes a
quantum phase transition from the quantum condensed phase to the
quantum conducting phase (see Fig. \ref{fig4}b). Here we assume
that the heterogeneous crystal is prepared in a state with $n=2$
LL collective phonons. In the quantum condensed phase these
phonons are localized by the impurity ion with no phonons at the
other sites. For sufficiently small $\alpha$ the adiabatic
condition requires a long interaction time and high value of
$\omega_{s}$. To avoid this problem we could increase $\alpha$ to
$0.1$. Then the quantum phase transition occurs approximately at
$\omega_{s}=2\pi\times0.4$ MHz and the adiabatic condition is
fulfilled within $60$ $\mu s$. The optical dipole trap can be
created by laser detuning up to $\Delta=-2\pi \times 300$ GHz red
of the resonance of the $\text{S}_{1/2}$ $\leftrightarrow$
$\text{P}_{1/2}$ transition of $^{43}\text{Ca}^{+}$. The frequency
$\omega_{s}$ can be achieved by laser power around $350$ mW with
waist radius of $w_{0}=5$ $\mu m$.

\section{Conclusions}

In conclusion we have presented a realistic scheme for observation
of phonon condensation and quantum phase transition of the
transverse local phonons in a heterogeneous ion crystal. Critical
behavior appears when the effective mass ratio is changed
continuously, which can be achieved by application of
light-induced dipole force. The features of the phase transition
are clearly visible even for a small number of ions, realizable
with the current ion trap technology. Impurity-doped ion crystals
offer the advantage that the laser light can be made to interact
only with the impurity ion. Moreover, the local phonon mode
readout of the impurity ion can be improved by using laser
frequencies selectively addressing its sidebands and internal
electronic states \cite{PAI}. The future extension of the proposed
technique may use multiple impurity ions for heat, excitation and
entanglement transport measurements, investigations of artificial
Josephson junctions and separation of phases.

We thank J.I. Cirac and A. Mering for discussions. This work has
been supported by the European Commission projects EMALI and
MICROTRAP, the Bulgarian NSF grants VU-I-301/07 and D002-90/08,
and the Elite programme of the Landesstiftung
Baden-W\"{u}rttemberg.


\end{document}